On the dual role of the *d*-electrons in iron-pnictides

Lev P. Gor'kov
*NHMFL, Florida State University, 1800 East Paul Dirac Drive, Tallahassee Florida 32310, USA
and L.D. Landau Institute for Theoretical Physics of the RAS, Chernogolovka 142432, Russia*

Gregory B. Teitel'baum
*E.K. Zavoiskii Institute for Technical Physics of the RAS, Kazan 420029, Russia*

Abstract

Recent Fe X-ray emission spectroscopy experiments [1] unveiled sizable local moments in iron-pnictides in the room-temperature paramagnetic state. In an effort to further clarify the notion of coexisting magnetic moments and itinerant carriers in iron-pnictides we focus on the interactions between the two subsystems. At a moderate on-site Coulomb repulsion the intra-atomic Hund's interaction leads to formation of the non-zero ("bare") local moments on the Fe-sites. We show that the Kondo-like exchange with the itinerant electrons may significantly renormalize the "bare" value of the moments manifested in different experiments. In turn, the itinerant carriers scatter on the renormalized moments that remain disordered in the paramagnetic phase. On the one hand, the scattering mechanism is responsible for high values of resistivity of the stoichiometric pnictides at the temperature of their transition into the antiferromagnetic phase, on the other; it washes out fine details of the Fermi surfaces. The results are rigorous and were obtained without use of any Born-type approximation. It also turned out that value of the local moment and the inverse free time for scattering of carriers on the moments tend in the limit of the strong Kondo exchange to the finite universal values. Independence of the results on the on-site Coulomb repulsion is then illustrated in frameworks of a simplistic model.

It is shown that the SDW transition is driven by the RKKY-interactions between the renormalized moments via exchange by the electrons –holes pairs. Applicability of the Boltzmann approach to transport in the multi-band pnictides is briefly discussed.

1. Introduction

Discovery in 2008 of superconductivity in doped ferro-pnictides [2] with temperatures of transition, $T_c$ up to 55 K [3] has opened new perspectives in condensed matter physics. Avalanche of subsequent publications confirmed emergence of the whole broad class of the Fe-based materials possessing many interesting properties, both in the normal and in the superconducting state. Encouraged by some similarities with the high temperature superconductivity (HTSC) in cuprates the experimental efforts were initially concentrated on search of the materials with even higher temperatures of the superconducting transition.

New materials are remarkable for the apparent competition between superconductivity and magnetism. They also reveal a rather uncommon behavior already at elevated temperatures. In what follows, we address some of these properties. To be more specific, we focus on the basic properties of local moments on the Fe-sites interacting with itinerant electrons in the paramagnetic phase. In addition, we study mechanisms that govern the structural and the magnetic instability in the stoichiometric materials. We also discuss the electrical transport in the same temperature interval.

The structural and the antiferromagnetic transitions in iron pnictides are close to each other and take place at temperatures $T_{SDW}$ that lie in the interval between 100 K and 200 K [3-5]. It is commonly accepted that the transition is driven by the magnetic instability, but is split into the two by the magneto-striction effects with the structural transition preceding [6].

In the literature there is no consensus yet concerning mechanisms that govern the phase transitions. Theoretical interpretations of the magnetic transition vary with the assumption about the nature of the paramagnetic phase. In the scenario of the itinerant electrons the magnetic transition could be driven by the spin density wave (SDW) instability coming from the so-called "nesting" of the Fermi surfaces. "Nesting" was very popular in the past at description of phase transitions in organic conductors. The nesting mechanism is the analogue to the Peierls transition in one dimension (1D) and proceeds from the instability of the supposedly congruent hole- and electron-like Fermi surfaces (FSs) at interactions [7]. Technically, the nesting model is rather flexible and is convenient for systematizing and interpreting even such delicate features in iron pnictides as the competition between SDW and superconductivity at doping [8, 9]. The fine details of the electronic spectrum in Fe-pnictides are usually studied in the band structure calculations and may depend on assumptions concerning the strength of the competing on-site interactions.

An alternative scenario of the antiferromagnetic transition is known as the $J_1$ –$J_2$ –model [10]. The antiferromagnetic ordering in [10] is ascribed to the superexchange mechanism characterized by competition of the two integrals, $J_1$ and $J_2$ for interactions between the local Fe moments on the nearest-neighbors (n.n) and the next- nearest neighbors (n.n.n.) sites, correspondingly.

Although FS nesting is compatible with a number of experimental observations for the evolution of spin excitations in electron/hole-doped iron-based superconductors, there are several aspects of the problem where such a scenario cannot be reconciled with experiments (See [11] for the review). Thus, the neutron scattering experiments reveal that fluctuating moments for $BaFe_2As_2$ and $BaFe_{1.9}Ni_{0.1}As_2$ are $m^2 = 3.17 \pm 0.16$ and $3.2 \pm 0.16$ $\mu_B^2$ per Fe(Ni), respectively. While these values are considerably smaller than those of the fully localized case, they are much larger than expected for value of the staggered magnetization in the nesting SDW scenario.

Recent Fe X-ray emission spectroscopy experiments unveiled the existence of sizable local moments in the room-temperature paramagnetic state [1] almost for all the pnictides studied. The observed local moments are similar in magnitude to those reported in the low-temperature neutron scattering experiments .

Vulnerability of the nesting scenario was emphasized already in the early review [12]. ARPES data [13] and more recent ARPES experiment [14], for instance, seem to show that the electronic spectrum of the antiferromagnetic phase of $BaFe_2As_2$ disagrees with the "nesting model".

The main unresolved issue concerns the role of electron-electron (*e-e*) interactions. The authors [15, 16] argued that electrons in the Fe-based materials are on the verge of localization in the Mott–Hubbard-type transition. Consequently, the electronic excitations cannot manifest same coherent character as in ordinary metals.

The view that neither localized nor itinerant approach can be fully correct for iron-pnictides was expressed by many other authors. However, as noticed, for instance, in [17], the direct Coulomb interactions seem to be moderate and, unlike [15, 16], correlations are driven by the Hund's rule coupling rather than by the on-site Hubbard repulsion.

These basic questions cannot be resolved without better understanding the nature of the paramagnetic phase. In the SDW-phase the values of the staggered magnetization turned out surprisingly small [18]. Elastic neutron scattering experiments are in favor of larger values of the local magnetic moments, as well [11, 19].

Theoretically the latter problem was addressed within the numerical approach that combines the density functional- and the dynamical mean field theory (DFT + DMFT) [20, 21]. It was demonstrated that, although the *d*-electrons possess the itinerant character, nevertheless, due to the strong on-site Hund's interactions they simultaneously contribute to the fluctuating magnetic moments on the Fe-sites.

As a whole, these results [20, 21] draw the picture of the "mixed valence" for the Fe-ions. The concept found the further development in the recent photoemission experiments [22] that can be summarized as follows. In the classical language, the *d*-electrons ever and again enter and leave the local

Fe- sites. For a short time while electrons are on the *d*-orbital, the strong local Hund's interaction tends to align their spins. These instantaneous values of the moment can be measured only by the "fast" experiments with a time scale ~ $10^{-16}$ sec [22]. (Other evidences in favor of such view were already known in the experimental literature. For the list of reference, see [22]).

Neither the direction, nor even the amplitude of the fluctuating moment is then fixed: locally, spin is not the good quantum number. In the language of quantum mechanics, the wave function of a single *d*-electron has both the local and the itinerant components. The interrelation between the components' amplitudes determines the average local moment. Thus, experiments [22] measure the probabilities of finding the electron in one or another state.

Accounting for such behavior of the *d*-electrons in all generality would present the formidable task for the theory. Most often than not, the studied phenomena are characterized by much longer time scales and the experiments measure on the Fe-sites some average spins value. Thermodynamics of the SDW transition gives one such example. The averaged local spins appear also in such kinetic characteristics as resistivity. Sizes of the electron-hole pockets in the BZ are small, and the Fermi velocities on single pockets are lower than the atomic value. This feature finds its reflection in the enlarged effective masses. In fact, the calculated band masses, $m_b$ usually equals to ~2-3 $m_e$ [18, 23]. The Hund's interaction increases these values further: $m^* \sim$ (3-4) $m_b$ [21]. Thereby, one may expect for the effective masses to be as large as $m^* \sim$ 6-8 $m_e$.

To the best of our knowledge, there was so far only one attempt to account for the dual character of the *d*-electrons theoretically. The authors [24, 25] considered the model in which, among all the *d*-bands some have the energy below the Fermi level; these electrons create the local magnetic moments on the Fe- sites. The rest form the bands of the itinerant electrons. The coexistence of the preformed moments and the itinerant electrons is postulated; the value of the local moments is fixed. The authors [24, 25] were interested in properties of the SDW phase and in a possibility of the superconducting pairing on the SDW background.

We consider the paramagnetic phase mainly of the so called "1111" and "122" Fe-pnictides. For the latter the staggered magnetic moments on the Fe-sites in the SDW state are low and may lie in the range of 0.2 - 1 $\mu_B$ (see [18]).

In all theoretical references cited before not much attention has been paid to properties of the itinerant electrons in the systems. Therefore, besides discussing instability of the paramagnetic phase itself, we stay on the resistivity behavior at temperatures above the transitions. The resistivity above $T_{SDW}$ is controlled by scattering of the electrons on the randomly oriented magnetic moments. Although the resistivity is metallic, $\rho(T)$ is large even in clean single crystals. Thus, in BaFe$_2$As$_2$ $\rho(T_{SDW}) \approx 0.25\ m\Omega \cdot cm$ [26]; it is even higher (at 200 K) ~0.5 $m\Omega \cdot cm$ in Sr122 [26, 27]. The authors [15, 16] argued that so high values signify violation of the Ioffe-Regel criterion. As distinct from [15, 16], we argue that to address this controversy one has to take into account the multi-band character of the energy spectrum of the Fe-pnictides. In addition, we compare our theory results with the resistivity data for LaOFeP [28, 29]. This "1111" compound is remarkable for absence of the SDW transition.

2. Interactions between local spins and itinerant electrons

The purpose of this Section is to study renormalization of the local moments in the presence of the Kondo-like exchange interactions with the itinerant sub-system. As point of departure, we assume single itinerant band of *d*-electrons with one electron- and one hole-like pocket.

The interaction between the local sites and the band electrons is chosen in the form:

$$\hat{H}_{int} = -\sum_i J S_i \hat{s}(R_i) \ . \qquad (1)$$

In Eq.(1) $\hat{s}(R_i)$ is the operator for the electronic spin of the unit cell located at the point $R_i$:

$$\hat{s}(R_i) = A_\square \hat{\psi}_\sigma^+(R_i) s_\sigma \hat{\psi}_\sigma(R_i) ,  \qquad (2)$$

where $A_\square \equiv a^2$ is the unit cell area. Each operator $\hat{\psi}$ (and $\hat{\psi}^+$) consists of the two components for the electronic and the hole pockets, $\hat{\psi}(R_i) = \hat{\psi}_e(R_i) + \hat{\psi}_h(R_i)$ correspondingly. For better transparency of arguments, we consider first interactions in the Hamiltonian (1) for only one itinerant pocket.

At the ferromagnetic sign, $J > 0$ Eq. (1) is the version of the Hund's interaction. Positive $J>0$ is therefore the natural choice in applications to the Fe-pnictides. (For a quantum spin, $\hat{S}_i$ and the antiferromagnetic sign of $J < 0$ the Hamiltonian Eq. (1) describes the Kondo problem). Following [22], the spin $S$ is *certain average* of the fluctuating non-conserving spin on Fe-sites. Correspondingly, we assume below that neither $S$ nor $s_\sigma$ are quantum operators. (The sign of $J$ is of no importance below). The sum in Eq. (2) is over the repeating spin index, $\sigma$ runs over the two directions of the electronic spin: $s_\sigma = \pm 1$.

To start with, calculate the correction second order in $J$, $\delta E_i^{(2)}$, to the energy of the center. It can be written as:

$$\delta E_i^{(2)} = J^2 S_i^2 \int dt' G(0;t-t') G(0;t'-t) . \qquad (3)$$

In Eq.(3) $G(0;t-t')$ stands for the Green functions, $G(R_i - R_i';t-t')$ at the coinciding spatial arguments: $R_i' \equiv R$. Like in the Kondo problem, the physics below is related to the properties of the Green functions at large time arguments. By definition [30]:

$$G_0(0;t) = -i\int [1-n(\varepsilon)]\nu(E_F)\exp(-i\varepsilon t)d\varepsilon ; \quad (t>0;\varepsilon>0) \qquad (4a)$$

and

$$G_0(0;t) = i\int n(\varepsilon)\nu(E_F)\exp(-i\varepsilon t)d\varepsilon ; \quad (t<0;\varepsilon<0) . \qquad (4b)$$

Here $n(\varepsilon)$ and $\nu(E_F)$ are the electrons' occupation number and the density of states at the Fermi level correspondingly. One easily finds from Eqs.(4a,b) the asymptotic behavior of the Green function at $|tE_F|>>1$:

$$G_0(0;t) \cong -\nu(E_F)[|t|-i\delta\,\text{sign}(t)]^{-1} . \qquad (5)$$

For classical spins, $S_i$ the calculations below become simpler in the frequency representation. For instance, the second order correction, Eq.(3) takes the form:

$$\delta E_i^{(2)} = J^2 S_i^2 \int \frac{d\omega}{2\pi} G_0^2(0;\omega) . \qquad (6)$$

In notations: $\xi = \varepsilon(p) - \mu$ ($\mu = p_F^2/2m \equiv E_F$), the Green function at the equal spatial arguments is [30]:

$$G_0(R_i = R_i; \omega) = \int \frac{A_\square d^2\vec{p}}{(2\pi)^2} G_0(p;\omega) \equiv \int \frac{A_\square d^2\vec{p}}{(2\pi)^2} \left\{ \frac{1}{\omega - \xi + i\pi\delta \, \text{sign}(\omega)} \right\} . \quad (7)$$

(Notice that from now on we include the factor $A_\square$ in the definition of the Green function, Eq.(7). In other words, the Green function at $R_i = R_i'$ is an "average" over the unit cell. Correspondingly, everywhere below $\nu(E_F)$ has the meaning of the density of states (DOS) per one spin direction and per one unit cell).

When substituting the identity: $\left\{ \frac{1}{\omega - \xi + i\pi\delta \, \text{sign}(\omega)} \right\} = P\left\{ \frac{1}{\omega - \xi} \right\} - i\pi \, \text{sign}(\omega)\delta(\omega - \xi)$ into Eq. (7) for the Green function, the first term can be omitted, because it produces no contribution from the vicinity of the chemical potential. Indeed, $P$ stands here for the principal value of the integral. For the band energy, $\varepsilon(p)$ were chosen symmetric respecting the chemical potential, $E_F$ such term would give zero *exactly*. Otherwise, the *P*- term leads to an insignificant spatial dependence in the Green function on the atomic scale. For this reason it is enough to leave under the integral sign in Eq. (7) only:

$$G_0(p;\omega) = -i\pi\nu(E_F)\text{sign}(\omega)\delta(\omega - \xi) . \quad (8)$$

We study the polaron-like effects in the system of spins on the local centers strongly coupled to spins of the itinerant carriers. Such physics has many formal similarities to the polaronic effect in a system with strong electron-phonon interactions. The latter has been studied recently in [31]. (The polaronic self-localization was discussed earlier for the A15 compounds in the Holstein model [32]). The mathematics is basically the same in both problems and below we merely list the main steps at the calculations.

For classical spins one can calculate all energy contributions from the interaction of Eq. (1) with the itinerant electrons by using the identity: $\partial E(S_1...S_N)/\partial J = \sum_{i,\alpha} S_i \hat{s}_\alpha(R_i)$. As in [31], equations for the Green function presented in the real space have the following form:

$$G_\sigma(R,R';\omega) = G_{0,\sigma}(R-R';\omega) + \sum_i G_{0,\sigma}(R-R_i;\omega) J\overline{S}^L_{i,\sigma}(\omega) G_{0,\sigma}(R_i - R';\omega)$$

$$+ \sum_{i \neq k} G_{0,\sigma}(R-R_i;\omega) J\overline{S}^L_{i,\sigma}(\omega) G_{0,\sigma}(R_i - R_k;\omega) J\overline{S}^L_{k,\sigma}(\omega) G_{0,\sigma}(R_k - R';\omega) + ... \quad (9)$$

(In accordance with notations in Eq. (2), in (9) we restored the dependence on the electronic spin indices).

The total *local* contribution into the center's energy from Eq. (9) has the following form:

$$\delta E_i = JS_i \int \frac{d\omega}{2\pi} G_0^2(0;\omega) J\overline{S}^L_i(\omega) . \quad (10)$$

In Eqs.(9, 10) at the expansion for spin $S_i(t)$ the summation over $J$ may be taken from *the left side*; with the help of Eq.(9) this will give the following self-evident equation for the $\overline{S}^L_i(\omega)$:

$$\overline{S}^L_{\sigma,i}(\omega) = S_i s_\sigma + S_i s_\sigma G_{0,\sigma}(R_i = R_i;\omega) J\overline{S}^L_{\sigma,i}(\omega) . \quad (11a)$$

The same procedure defines the variable $\overline{S}^R_j(\omega)$:

$$\overline{S}^R_{\sigma,j}(\omega) = S_j s_\sigma + J\overline{S}^R_{\sigma,j}(\omega)G_{0,\sigma}(R_j = R_j; -\omega)S_j s_\sigma, \quad (11b)$$

if summation over $J$ were performed from *the right side*. $\overline{S}^R_j(\omega)$ and $\overline{S}^L_j(\omega)$ are complex conjugate.

It must be emphasized that at derivation of Eqs.(11a,b) no smallness of the interactions $J$ in the Hamiltonian (1) was assumed. Omitting details, the total contribution into the local energy of one center is:

$$\delta E_i = -(W/2\pi)\ln\{1 + [\pi\nu(E_F)JS_i]^2\}. \quad (12)$$

($W$ is the band width). Other terms in Eq.(9) lead to contributions into the total energy from the interactions between different centers. As it will be shown, relevant to the physics of Fe-pnictides are the RKKY-type interactions via exchange by the electron-hole pairs. The latter are investigated in the two-pocket model below.

Consider now the itinerant electrons. Scattering of electrons on the local moments above $T_{\text{SDW}}$ is given by the average of $J\overline{S}^L_{\sigma,i}$ that stands in the denominator of the Green function:
$G(p;\omega) = [\omega - \xi + <J\overline{S}^L>]^{-1}$. From Eq. (11a):

$$J\overline{S}^L(\omega) = JS[1 - i\,\text{sign}(\omega)\pi\nu(E_F)JS]^{-1}. \quad (11')$$

Above SDW transition the onsite Fe-spins $S_i$ are not oriented and for $<J\overline{S}^L_i>$ it follows:

$$<J\overline{S}^L_i> = i\,\text{sign}(\omega)\pi\nu(E_F)(JS)^2 / [1 + (\pi\nu(E_F)JS)^2]. \quad (13)$$

Using again Eq.(11') in the expression for the *correlator* $<S_i(0)S_i(t)>$ one arrives back to the notion of the randomly oriented *renormalized* spins:

$$<S_i(0)S_i(t)> = \int\frac{d\omega}{2\pi}<S\overline{S}^L_i(\omega)>e^{i\omega t} \equiv \delta(t)\overline{S_i^2}, \quad (14)$$

where the average square of renormalized moment is

$$\overline{S_i^2} = \frac{\pi\nu(E_F)JS^2}{1 + [\pi\nu(E_F)JS]^2}. \quad (14')$$

(In the paramagnetic phase at averaging in Eqs.(13, 14) the terms odd in $S$ dropped out).

At large $J$ one has: $\overline{S_i^2} = 1/(\pi\nu(E_F)J)$.

The itinerant carriers form a "cloud" at a site (*i*). The average value of its spin is:

$$<s_i> = \int\frac{d\omega}{\pi}G_0^2(0,\omega)J\overline{S}^L_i(\omega) = JSW\pi\nu^2(E_F)/[1 + (\pi\nu(E_F)JS)^2]. \quad (14'')$$

3. Multi-band case

Extension of the above results to the case of one electronic- and one hole- pockets is trivial. The Hamiltonian of Eq. (1) would now include interactions with carriers from both pockets, including the non-diagonal terms. Correspondingly, in its general form the Hamiltonian is:

$$\hat{H}_{int} = -\lambda \sum_{i;l,t} S_i J_{lt} \hat{s}_{lt}(R_i), \qquad (1')$$

where:

$$\hat{s}_{lt}(R_i) = A_{\square} \hat{\psi}^+_{l,\sigma}(R_i) s_\sigma \hat{\psi}_{t,\sigma}(R_i). \qquad (2')$$

The indices ($l,t$) stand for the electron and the hole pockets. (As usual, the summation runs over the repeated Latin indices).

It is convenient to introduce one common parameter $\lambda$ in $\partial E(S_1...S_N)/\partial \lambda = \sum_{i,lt} S_i J_{lt} \hat{s}_{lt}(R_i)$ ($\lambda \in \{0,1\}$). Eq.(9) for the Green functions preserves the same structure, except that the Green function now depends on $\lambda$ and has the matrix form: $G^{(\lambda)}_{lt;\sigma}(R,R';\omega)$. The free Green function $G_{0,\sigma;lt}(R,R';\omega)$ is the diagonal matrix:

$$G_{0,\sigma;lt}(R,R';\omega) \Rightarrow \begin{pmatrix} G^{(e)}_{0,\sigma}(...) & 0 \\ 0 & G^{(h)}_{0,\sigma}(...) \end{pmatrix}. \qquad (15)$$

Instead of Eq. (11) for the single center, one has to calculate $J_{lt} G_{lt}(0;\omega)$ where $G_{lt}(0;\omega)$ obeys the equation:

$$G_{\sigma;lt}(0;\omega) = G_{0,\sigma;lt}(0;\omega) + G_{0,\sigma;lm}(0;\omega) S_i J_{mp}(\omega) G_{pt}(0;\omega). \qquad (16)$$

While the matrix form of Eqs. (15, 16) significantly complicate calculations of the total energy the physics eventually suffers only minor changes.

4. SDW instability

Turn now to the antiferromagnetic (SDW) transition that is driven by the RKKY interaction. In case of the stoichiometric Fe-pnictides the structural vector $\vec{Q}$ of the SDW phase is commensurate and connects centers of the hole-like pocket at the $\Gamma$-point and the electron pocket at the X-points (in the unfolded Brilloin zone (BZ) is commensurate). Thereby, the transition is controlled by the interaction $J_{eh}$ between carriers from the two pockets. To demonstrate the basic physics, we restrict ourselves by the model of the preceding Sect. 3 with the two bands only and assume that the non-diagonal exchange $J_{eh}$ is relatively small: $J_{eh} \ll J_e, J_h$. The RKKY interaction acquires the form:

$$E_{i,k} \equiv E(R_i, R_k) = -i 2 S_i S_k J_{eh}^2 \Pi_{eh} \int (d\omega/2\pi) G_{0,\sigma;e}(R_i - R_k;\omega) G_{0,\sigma;h}(R_i - R_k;\omega). \qquad (17)$$

The factor $\Pi_{eh}$ stands for the renormalized value of the spins $S_i, S_j$. Making use of Eqs. (11a,b) one finds that the product:

$$\Pi_{eh} \equiv |\bar{S}_i^L(\omega)/S_i|^2 |\bar{S}_j^L(\omega)/S_j|^2 = [1+(\pi\nu_e(E_F)J_e S)^2]^{-1}[1+(\pi\nu_h(E_F)J_h S)^2]^{-1} \quad (17')$$

is frequency independent. Comparison between Eq.(17') and Eqs.(13,14') shows that the renormalization factors may differ depending on the effect under consideration and on the structure of the electronic spectrum.

In the diagrammatic representation at finite temperatures [30], Eq. (17) is:

$$E_{i,k} = -4 S_i S_k J_{eh}^2 \Pi_{eh} T\pi \sum_{\omega_n} \int \frac{A_\Box d\vec{p}}{(2\pi)^2} G_{0,h}(\varepsilon(\vec{p});\omega_n) G_{0,e}(\varepsilon(\vec{p}-\vec{Q});\omega_n). \quad (18)$$

(Additional factor 2 comes from two spin directions). It is convenient to introduce the new notation:

$$E_{i,k} \equiv S_i S_k J_{eh}^2 \Pi_{eh} \pi\nu(E_F) F(T,\eta). \quad (19)$$

After summation over the thermodynamic frequencies, $\omega_n = (2n+1)\pi T$ in (18) one arrives to the following well known expression:

$$F(T,\eta) = \int \frac{\Omega d\vec{p}}{(2\pi)^2 \nu(E_F)} \left( \frac{1}{\varepsilon_h(\vec{p}) - \varepsilon_e(\vec{p}-\vec{Q})} \right) \left\{ \tanh\left(\frac{\varepsilon_h(\vec{p})}{2T}\right) - \tanh\left(\frac{\varepsilon_e(\vec{p}-\vec{Q})}{2T}\right) \right\}. \quad (20)$$

(Below the parameter $\eta$ will characterize the dependence of $F(T,\eta)$ Eq. (20) on such factors as the anisotropy or doping. The integration in (20) extends over the energy interval $\tilde{W}_{eh}$ that corresponds to the overlap of the electron- and hole-like bands).

No rigorous solutions for the antiferromagnetic transition are known and one usually resorts to the mean-field approach. The method consists in analyzing the linear response of the system to the external "staggered" field, $B_{ext}^{stag} \Rightarrow B_{ext}(Q)$.

In the mean-field approach the system is unstable with respect to the magnetic transition when the following criterion is fulfilled:

$$1 = \frac{J_{eh}^2 S^2 \Pi_{eh} \pi\nu(E_F)}{T_{SDW}} F(T_{SDW},\eta). \quad (21)$$

Here $F(T_{SDW},\eta)$ plays a secondary role as a coefficient in front of the $1/T$ prefactor. Its magnitude depends on fine details of the *e-h* spectrum, i.e., on the anisotropy or doping. Generally, we do not specify the band spectrum, but such factor would increase due to the logarithmic singularity at the wave vectors coinciding with the nesting vector $Q$, if any. Experimentally, in the stoichiometric Fe-pnictides $Q$ is commensurate, but may deviate from the commensurability at a large enough doping level.

The choice of the overlap $\tilde{W}_{eh}$ as the parameter is not convenient in what follows. A more familiar energy scale can be defined by considering first $F(T,\eta)$, Eq. (19), in the approximation of the *exact nesting*:

$$F(T,0) = \int \frac{A_\Box d\vec{p}}{(2\pi)^2 \nu(E_F)} \left( \frac{th(\xi/2T)}{\xi} \right) = 2\ln\left(\frac{\gamma \tilde{W}_{eh}}{\pi T}\right), \quad (22)$$

Let $T_{SDW}^{(0)}$ be the temperature of the SDW transition for this case:

$$T_{SDW}^{(0)} = 2J_{eh}^2 S^2 \pi \nu(E_F) \ln(\frac{\gamma \bar{W}_{eh}}{\pi T_{SDW}^{(0)}}) \quad . \tag{23}$$

Considering doping or the anisotropy recall that, realistically, the latter always are large enough to cut the logarithmic singularity in (22); then $\ln(\gamma \tilde{W}_{eh} / \pi T) \Rightarrow \ln(\gamma \tilde{W}_{eh} / \eta)$ and from (21) for the dependence $T_{SDW}(\eta)$ it follows:

$$T_{SDW} = T_{SDW}^{(0)} - 2J_{eh}^2 S^2 \pi \nu(E_F) \int \ln\left[\frac{\eta(\vec{k})}{\pi T_{SDW}^{(0)}}\right] \frac{d\Omega_{\vec{k}}}{2\pi} \quad . \tag{24}$$

(Integration in Eq. (24) averages over the Fermi surface the anisotropy in the parameter $\eta$, if any).
So far, in the discussion above we fully ignored the electron-electron interactions. The latter can be of interest, especially at calculations of $F(T,\eta)$ in the logarithmic approximation. The corresponding analysis would lead to the expressions for $F(T,\eta)$ that are familiar from numerous publications on the "nesting" scenario. (See, e.g., in [33]). It does not change the general structure of Eq. (21) and we do not stay on further details here.
Eq. (21) reveals the main difference from the "nesting" scenario. In our case the interaction takes place between the nonzero (renormalized) moments on different sites that exist already in the paramagnetic phase, while in the nesting scenario the finite magnetization appears only below $T_{SDW}$. We repeat once again that the electronic spectrum in Eq. (20) has a general character. Even more, the method of Eq.(21) and the analytic $T_{SDW}(\eta)$ - dependence in Eq. (24) differ from that one in the "nesting" scenario. The "nesting" scenario relies on the "congruency" of the electron and holes pockets that results in the logarithmic singularity in the $F(T_{SDW},\eta)$-function that is compensated then by the interaction constant [34]. (Besides, note that Eq. (24) involves the independent parameter, $J_{eh}$).

5. Limit of strong interactions

From Eq.(11') we derived the expression for the imaginary part of the self energy Eq.(13) that determines the inverse mean time for scattering of electrons on the disordered moments in the paramagnetic phase:

$$\frac{\hbar}{\tau} = \frac{\pi \nu(E_F) J^2 S^2}{1 + [\pi \nu(E_F) JS]^2} \quad . \tag{25}$$

The combination $JS$ is the product of the coupling strength and the "bare" spin. It is remarkable that in the limit of large $J$ this expression acquires a very simple form:

$$\hbar / \tau = 1 / \pi \nu(E_F), \tag{25'}$$

Thereby, any information about the local spins on the Fe-site drops out from Eq. (25').

That prompts the question whether these results could be correct even in the extreme limit of the Coulomb repulsion (or the Hubbard's $U$) on a Fe-site being much larger than the *intra-atomic* Hund's exchange. In other words, the competition between the *on-site* Coulomb and Hund's interactions becomes replaced by the competition between strong local Hubbard repulsion $U$ and the strong Kondo exchange with the itinerant carriers Eq. (1). For brevity, consider the simplistic model of a single itinerant $d$-band with one electron- and one hole-like pocket centered at different points in the BZ and the Hubbard $U$ strong enough to forbid the double on-site occupancy (the "$U$-model").

In the $U$-model $E_i$ to build up the on-site magnetic moment $S_i$ would cost the energy [35]:

$$E_i = U \frac{S_i^2}{2}. \qquad (26)$$

(In the above $S$ could be considered either as the spin ($S>1$) or as the magnetic moment measured in units of the Bohr magneton (see also [35])).

Return now to calculation of the second order correction, $\delta E_i^{(2)}$ Eq. (6). Substituting Eq. (8) for the Green functions and performing the trivial integration, one arrives to:

$$\delta E_i^{(2)} = -(W/2\pi) J^2 S_i^2 [\pi \nu(E_F)]^2 . \qquad (27)$$

(Here again $W$ is the bandwidth, $\nu(E_F)$ stands for the density of states (DOS) per unit cell per single spin. For a parabolic two-dimensional (2D) spectrum DOS does not depend on $E_F$).

The negative sign in (27) shows that the non-magnetic state of the center at large enough coupling constant may become unstable. In that case the electron- ion interactions, Eq. (1) will build up the local moments on the each site $R_i$. The threshold value, $J_0$ for onset of the instability is defined by equating (27) to the energy of the "bare" center, Eq. (26):

$$U \frac{S_i^2}{2} = (W/2\pi) J^2 S_i^2 [\pi \nu(E_F)]^2 . \qquad (28)$$

Introduce the notation: $J^2 = J_0^2 \Lambda^2$. From Eq.(28) for $J_0$ one finds:

$$J_0^2 = U \frac{1}{\pi W \nu(E_F)^2} . \qquad (28a)$$

The total energy is the sum of Eqs. (12, 26):

$$E_{i,tot} = U \frac{S_i^2}{2} - \frac{W}{2\pi} \ln[1 + \pi^2 \nu(E_F)^2 J_0^2 \Lambda^2 S_i^2]. \qquad (29)$$

Denote also:

$$\tilde{S}_i^2 = [\pi J_0 \nu(E_F)]^2 S_i^2 . \qquad (28b)$$

In the new notations Eq. (29):

$$E_{i,tot} = \left(\frac{W}{2\pi}\right) \{\tilde{S}_i^2 - \ln[1 + \Lambda^2 \tilde{S}_i^2]\} . \qquad (30)$$

At $\Lambda^2 > 1$ $E_{i,tot}$ in Eq. (30) has minima at $\tilde{S}_\pm = \pm\sqrt{1 - \Lambda^{-2}}$ .

Thereby, the Kondo exchange induces the on the site $(i)$ the non-zero spin:

$$S_i^2 = (\frac{W}{U\pi})(1-\Lambda^{-2}). \quad (31)$$

The energy at the minimum:

$$E_{i,tot}^{min} = (W/2\pi)[1-\Lambda^{-2} - \ln\Lambda^2]. \quad (32)$$

At small $\Lambda^2 \ll 1$ Eq. (30) is the potential energy of an anharmonic oscillator:

$$E_{i,tot} = \left(\frac{W}{2\pi}\right)\{\tilde{S}_i^2(1-\Lambda^2) + (\frac{1}{2})\Lambda^4 \tilde{S}_i^4\}. \quad (33)$$

All results of Sect. 2 apply once the Ising spins are formed at $\Lambda^2 > 1$. In particular, in the paramagnetic phase the inverse mean free time for scattering of electron on disordered spins and the average square of the local moment are given by the same Eqs.(13, 14'). Thus:

$$\overline{S_i^2} = \frac{\pi\nu(E_F)JS^2}{1+[\pi\nu(E_F)JS]^2}. \quad (34)$$

Hence, again $\overline{S_i^2} = 1/(\pi\nu(E_F)J)$ at large $J$. Actually, according to (28b), the limit of $[\pi\nu(E_F)JS]^2 \gg 1$ is the limit of large $\Lambda^2 \gg 1$: $[\pi\nu(E_F)JS]^2 \equiv \Lambda^2 \tilde{S}^2 \gg 1$.

Returning to the case of the two pockets, the stability condition (see Eq. (27)) acquires the form:

$$U\frac{S_i^2}{2} = S_i^2\{(W_{ee}/2\pi)J_{ee}^2[\pi\nu_e(E_F)]^2$$

$$+2(W_{eh}/2\pi)J_{eh}^2[\pi^2\nu_e(E_F)\nu_h(E_F)] + (W_{hh}/2\pi)J_{hh}^2[\pi\nu_h(E_F)]^2\}. \quad (35)$$

In Eq. (35) $W_{ee}$ and $W_{hh}$ are the bandwidths for the electron- and the hole-like bands, correspondingly. As to $W_{eh}$, it was already defined above as the energy difference between the top and the bottom of the $h$-and the $e$- bands. (The chemical potential must be inside each of the two bands).
 Again, Eq. (35) demonstrates that provided the strength of interactions exceeds some threshold the system becomes unstable for the multi-band case as well. At non-zero $J_{eh}$ the subsequent calculation, however, become cumbersome; we will not dwell upon such case further.
 Same calculations as above show that at $\Lambda^2 > 1$ the SDW instability corresponds to the transition in the system of the interacting Ising spins. Then $\tilde{S}^2 = (1-\Lambda^{-2})$ and Eq. (21) becomes:

$$T_{SDW} = \left(\frac{\Pi_{eh}}{\pi\nu(E_F)}\right)\left(\frac{\Lambda^2-1}{\Lambda^2}\right)\left(\frac{J_{eh}^2}{J_0^2}\right)F(T_{SDW},\eta). \quad (21')$$

For completeness, consider also the case of small $\Lambda^2 < 1$. Above $T_{SDW}$ local moments appear only due to the thermal fluctuations:

$$<<\tilde{S}^2>> = \{\int \exp(-E_{tot}/T)\tilde{S}^2 d\tilde{S}\} / \int \exp(-E_{tot}/T)d\tilde{S} . \qquad (36)$$

Analytically, the average (36) can be calculated only at $\Lambda^2 \ll 1$. Substitution of Eq. (33) for $E_{tot}$ gives:

$$<<\tilde{S}^2>> \approx (\frac{2\pi T}{W})\{1 - \frac{9}{16}\Lambda^4(\frac{2\pi T}{W})\} . \qquad (36')$$

The second term in the brackets at $T = T_{SDW}$ is small and instead Eq. (21') one has:

$$\frac{1}{\pi \nu(E_F)}(\frac{J_{eh}^2}{J_0^2})F(T_{SDW},\eta) = \frac{W}{2\pi} . \qquad (37)$$

In Eq. (37) $W \sim 1 eV$, and one may immediately realize that for a solution for $T_{SDW}$ to exist on same scale of $100K \div 200K$ an enhancement should occur in the left side of Eq. (37) at $T \sim T_{SDW}$. "Nesting" features in the electronic spectrum under the integral in (20) for $F(T,\eta)$ could help.

However, since experimentally in stoichiometric pnictides the SDW transition always takes place via ordering of the local moments, we do not dwell any further upon details of the SDW transition described by Eq. (37).

We conclude the discussion of the simple model Eq. (26) by calculating the inverse mean free time for scattering of electrons on the thermally fluctuating local moments at $\Lambda^2 < 1$:

$$\hbar/\tau = (2\Lambda^2/W\nu(E_F))T . \qquad (38)$$

(Actually, as in (36'), $\hbar/\tau$ is calculated with $\Lambda^2 \ll 1$ with small terms omitted).

6. Resistivity

Rough estimates of the mean free path, $\ell$ in [15, 16] from data on the resistivity of polycrystalline samples of LaFeAsO led the authors to the conclusion that conductivity in the material cannot be described within the standard Boltzmann equation: according to their analysis $k_F \ell \sim 1$, where $k_F$ stands for the Fermi momentum. First, the measurements on the polycrystalline samples do not provide the intrinsic value of resistivity. In addition, the multi-band character of the electronic spectrum in iron-pnictides was not taken into account. Below we repeat the analysis [15, 16] for a few iron pnictides.

Data for the resistivity, $\rho(T)$ at elevated temperatures from single crystals are scarce. For $BaFe_2As_2$ we use the single-crystal data [26]. Unlike [15, 16], whenever it is possible, we analyze the contributions into conductivity from each single pocket separately:

$$\sigma_i = n_i e^2 \tau_i / m_{eff,i} . \qquad (39)$$

It is instructive to start with the universal form of the mean free time, Eqs. (14, 14') for each pocket:

$$\hbar/\tau_i = 1/\pi\nu_i(E_F) . \qquad (40)$$

(Recall that in the limit of strong interaction Eq. (40) is the same in both models).

Then

$$\tau_i = \pi \nu_i(E_F)\hbar = \frac{2\pi^2 \hbar (1/2) m_{eff,i} a^2}{(2\pi\hbar)^2} = \frac{a^2 m_{eff,i}}{2\hbar}. \qquad (41)$$

Correspondingly,

$$\sigma_i = (e^2 \pi / 4c\hbar)(ap_{F,i}/\pi)^2 \approx (ap_{F,i}/\pi)^2 1.2 \cdot 10^{15} \text{ sec}^{-1}. \qquad (39')$$

(Note that in this limit the masses drop out from the expression for conductivity).

BaFe$_2$As$_2$. There are two FeAs- layers in "122" per unit cell. Taking the sum of the values of $(ap_{F,i}/\pi)^2$ [23] over all four 2D- pockets, for resistivity in BaFe$_2$As$_2$ at $T=T_{SDW}$ one obtains $\rho(T_{SDW}) \approx 1 m\Omega \cdot cm$. Experimentally, [26, 36] $\rho(T_{SDW}) \approx 0.14 m\Omega \cdot cm$. In the $U$- model this difference could be compensated, in principle, by an appropriate choice of $\Lambda^2$ in Eq. (31). Still, the difference is too big and the attempt to compensate it by a proper value of $\Lambda^2$ would pose BaFe$_2$As$_2$ too close to the threshold Eq. (27). No matter how, it is obvious that the "universal" expression Eq. (40) significantly underestimate the conductivity in "122". Hence, in the scenario of the preformed local moments the Hund's interaction, Eq. (1) in the system must not be too strong.

"1111". Repeating the same analysis for a few "1111" compounds [23] in different compounds gave the values $\rho(T_{SDW})$ that usually lie between 1 and $1.16 m\Omega \cdot cm$. This is in a rather good agreement with the experiments as discussed in [18, 37, and 38].

With the notion [22] of rapidly fluctuating moments on the Fe-sites the question arises whether or not the kinetic characteristics, such as conductivity, can be analyzed in terms of the same "average" spins, as it was done above for the SDW transition.

In the limit of strong interaction, Eq. (41) the characteristic time is proportional to the effective mass on the pocket. Even with the mass, $m_{eff,i}$ enhanced up to 6 $m_e$ [21, 23] and $a \approx 4 \cdot 10^{-8} cm$ one obtains for $\tau_i \sim 0.5 \cdot 10^{-14}$ sec $<< 10^{-16}$ sec. Depending on the mass, the energy scale is $\hbar/\tau \approx 0.13 - 0.4 \, eV$ and is comparable with the corresponding Fermi energies (typically of the order of ~ few tenths of $eV$ [18, 23]).

One concludes that the analysis with the use of Eqs. (40, 41) for $\tau_i$, will give for iron pnictides the results that seem to be close to the border of applicability of the standard Boltzmann expression. Fortunately, as it was noticed above, for BaFe$_2$As$_2$ Eqs. (40, 41) overestimate its resistivity at and above $T_{SDW}$.

Taking seriously the factor ten differences between the calculated, $\rho(T_{SDW}) \approx 1 m\Omega \cdot cm$ and the experimental value, $\rho(T_{SDW}) \approx 0.14 m\Omega \cdot cm$ [36] one may try to estimate the Hund's interaction coupling. Comparison of the expressions (40) and (14') leads to $(\pi\nu(E_F)JS)^2 \approx 0.1$. With $1/\pi\nu(E_F) \approx 0.13 - 0.4 eV$ it gives the surprisingly low values for $JS$ to be between 0.04 and 0.13 $eV$. With $S \sim 1-2$ from the photoemission experiment [22] such estimate gives for the Hund's constants too low a value which seems to be in conflict with the commonly accepted values of the Hund's integral of order of 0.2-0.35 $eV$ [20, 39]. It is the time to remember that the measured staggered SDW magnetic moments are between 0.2 and 1 $\mu_B$ [18]. Therefore, it seems that the rapidly fluctuating moments on the Fe-sites contribute into the transport characteristics only through the time- averages, similar in value to those ones observed in the SDW phase.

We conclude this discussion by the comment on LaOFeP. The SDW transition is absent in the latter. Whether large moments on the Fe-sites are present in this material is not known. Our model, however, provides, assuming $\Lambda^2 \ll 1$, the tentative explanation for the large coefficient in front of the linear in T dependence of resistivity of this material above ~20 K [28, 29]. With Eq.(38), $\hbar/\tau_i = (2\Lambda^2/W_i \nu_i(E_F))T$ one can make a rough estimate choosing for each pocket the same bandwidth, $W_i \sim 1eV$ and taking for simplicity $\Lambda^2 = 1/2$; this would give for the resistivity $\rho \sim 1.6 \cdot 10^{-19} \cdot T$ sec $\sim 1.6 \cdot 10^{-4} \cdot T$ $m\Omega cm$, rather close to the experimental data [28, 29].

6. Conclusion

In summary, we studied the dual role of the *d*-electrons in the stoichiometric Fe-pnictides in the paramagnetic phase in the model of the local moments and the itinerant electrons interacting via the Kondo-type Hamiltonian. At a moderate on-site Coulomb interaction the "bare" local moments become renormalized via the Kondo-like exchange interactions with itinerant electrons. In turn, the itinerant carriers scatter of on the renormalized moments that are disordered in the paramagnetic state. The latter mechanism explains the surprisingly high values of resistivity of the stoichiometric pnictides at the temperature of their transition into the antiferromagnetic phase.

It was shown that instability to the onset of the SDW order is caused by the RKKY-interactions between the *renormalized* local moments via exchange by the electron –hole pairs; in case of stoichiometric Fe-pnictides the structural vector of the SDW phase is pre-determined by the positions of the electron- and the hole-pockets in the BZ. The dependence of the transition temperature, $T_{SDW}$ on anisotropy and doping is found different from that one in the "nesting" scenario.

Generally, the assumed *intra-atomic* Hund's exchange is comparable or can even prevail over the Coulomb onsite repulsion. However, in the limit of strong Kondo-like exchange between the local moment and itinerant carriers the value of the local moment itself and the effective mean free time for carriers scattered on the moments tend to the finite values. The independence of the result on the strength of the on-site Coulomb repulsion was then justified by solving a simplistic model.

At calculations of conductivity of iron-pnictides in the paramagnetic phase the multi-band character of the electronic spectrum in iron pnictides had to be taking into account. Contributions into the total conductivity from several electron- and hole-like pockets were then considered one by one.

For the "1111" group even simplistic model leads to the experimental values of resistivity at $T_{SDW}$. In case of the "122" iron-pnictides the limit of strong interactions overestimates the resistivity value at $T_{SDW}$. Unlike the "1111" pnictides, it is found that the Born approximation already provides the satisfactory results for resistivity of $BaFe_2As_2$ thus confirming applicability of the standard kinetic analysis.

The linear in *T* dependence of resistivity in LaOFeP was explained in terms of the thermal fluctuations of local spins. Unlike other "1111" iron pnictides, the SDW transition is not observed in LaOFeP.

All results in the paper were obtained owing to use of the new exact equations for the electronic Green functions in the real space.

7. Acknowledgements

The authors are grateful to I. A. Nekrasov and M. V. Sadovskii for sharing their unpublished data on the band parameters for many iron pnictides compounds. The work of L.P.G. was supported by the NHMFL through NSF Grant No. DMR-0654118 and the State of Florida, that of G.B.T. through the RFBR under Grant No. 10-02-01056.